\documentclass[3p]{elsarticle}

\usepackage[cmex10]{amsmath, mathtools}
\usepackage{amsfonts, amssymb, amsthm}
\usepackage{algorithmic}
\usepackage{algorithm}
\usepackage{array}
\usepackage[usenames,dvipsnames]{color}
\usepackage[caption=false,font=footnotesize]{subfig}
\usepackage{setspace}

\theoremstyle{plain}
\newtheorem{thm}{Theorem}[section]

\theoremstyle{definition}
\newtheorem{defn}[thm]{Definition}
\newtheorem{lem}[thm]{Lemma}

\newtheorem{rem}[thm]{Remark}

\newtheorem{conj}[thm]{Conjecture}

\newtheorem{find}[thm]{Finding}

\newcommand{\col}{\mathcal}
\newcommand{\dom}{\mathbb}

\begin{document}

\begin{frontmatter}

\title{Discretization of Planar Geometric Cover Problems}

\author[addr1]{Dae-Sung Jang}
\ead{dsjang@lics.kaist.ac.kr}
\author[addr1]{Han-Lim Choi\corref{cor1}}
\ead{hanlimc@kaist.ac.kr}
\cortext[cor1]{Corresponding author.}
\address[addr1]{Division of Aerospace Engineering, KAIST, Daejeon, 305-701, Republic of Korea}

\begin{abstract}
We consider discretization of the `geometric cover problem' in the plane:
Given a set $P$ of $n$ points in the plane and a compact planar object $T_0$, find a minimum cardinality collection of planar translates of $T_0$ such that the union of the translates in the collection contains all the points in $P$.
We show that the geometric cover problem can be converted to a form of the geometric set cover, which has a given finite-size collection of translates rather than the infinite continuous solution space of the former.
We propose a reduced finite solution space that consists of distinct canonical translates and present polynomial algorithms to find the reduce solution space for disks, convex/non-convex polygons (including holes), and planar objects consisting of finite Jordan curves.
\end{abstract}


\end{frontmatter}

\section{Introduction}\label{sec:intro}

A {\it planar translate} or {\it translate} $T_i$ of an object $T_0$ in the plane is an object created by a planar translation of $T_0$, which is termed the {\it prototype} of $T_i$ herein.
Given a set $P$ of $n$ points in the plane and a compact planar object $T_0$, the {\it geometric cover} (or covering) {\it problem} in the plane is to find a minimum cardinality collection $\col{T}^*$ of planar translates of $T_0$ such that the union of the translates in $\col{T}^*$ covers all the points in $P$.
The geometric cover problems for identical axis-parallel squares and unit disks are proven to be NP-hard \cite{Meg84}, and Hochbaum and Maass \cite{Hoc85}, and Gonzalez \cite{Gon91} devised PTASs for the problems in a multi-dimensional space.

A very closely related problem to the geometric cover in a discrete and finite domain is the {\it geometric set cover}, which is to find a minimum cardinality subcollection $\col{T}^*\subseteq\col{T}$, covering all the points, in a given collection $\col{T}$ of $m$ planar translates of $T_0$.
The general set cover problem is NP-hard and has no polynomial algorithm with an approximation ratio less than $c\log n$ \cite{Raz97}, whereas the geometric set cover is approximable with a lower approximation ratio by the fact that many geometric objects have set systems of finite VC-dimension which guarantee the existence of $\epsilon$-net \cite{Hau87}.
Br\"{o}nnimann and Goodrich \cite{Bro95} addressed the first $O(\log|\col{T}^*|)$-approximation algorithm for finite VC-dimensional set systems, and an $O(|\col{T}^*|)$-approximation algorithm for disks in the plane.
Clarkson and Varadarajan \cite{Cla07} gave an $O(\log|\col{T}^*|)$-approximation algorithm for fat triangles, pseudo-disks, a family of fat objects, and a constant factor approximation algorithm for similar-sized fat triangles and fat objects, fat wedges, and unit cubes in $\dom{R}^3$.
Laue \cite{Lau08} also presented a constant factor approximation algorithm for polytopes in $\dom{R}^3$.

The most actively studied topic in the geometric set cover is the {\it unit disk cover}, which is known to be NP-hard in the strong sense \cite{Fow81,Joh82}.
A series of studies on constant factor approximation of the unit disk cover after the first elaboration \cite{Bro95} is as follows:
108-approximation algorithm by C\u{a}linescu et al. \cite{Cal04}, %(102$\rightarrow$108: corrected by Narayanappa)
72-factor by Narayanappa and Voj\v{t}echovsk\'{y} \cite{Nar06},
38-factor by Carmi et al. \cite{Car07},
22-factor by Claude et al. \cite{Cla10},
18-factor by Das et al. \cite{Das11},
15-factor by Fraser and L\'{o}pez-Ortiz \cite{Fra12},
and recently, (9+$\epsilon$)-factor by Acharyya et al. \cite{Ach13}.
This endeavor to constant factor approximation algorithms has partially arisen from the lack of practicality of PTAS \cite{Ach13,Das11}.
The first PTAS for the unit disk cover was presented by Mustafa and Ray \cite{Mus09} which runs in $O(m^{2(\frac{8\sqrt{2}}{\epsilon})^2+1}n)$ time for $0<\epsilon\leq2$, and another one devised by Liao and Hu \cite{Lia10} takes $O(nm^{O(\frac{1}{\epsilon^2}\log^2\frac{1}{\epsilon})})$ time, where $n=|P|$ and $m=|\col{T}|$.

%{\red a PTAS for axis parallel unit squares, Erlebach and Leeuwen: }

In spite of the attention to the {\it discrete} geometric set cover, the geometric cover problem with infinite possible locations of translates over a continuous domain has been covered infrequently;
in the above literature, it is sometimes confused with the geometric set cover and other times it is regarded as a somewhat different matter.
Although the relation between the two problems has often been overlooked, it can be expected that a close relationship is derived from the identification of an instance of the geometric cover problem in discrete form.
If there exists some efficient method to extract all possible combinations of points in $P$ that can be covered by a translate from infinite possibilities and then we can build the collection $\col{T}$ of translates to convert the instance of the geometric cover problem to an instance of the geometric set cover.
This gives rise to a conjecture that the geometric cover problem is a special case of the geometric set cover.
However, any algorithm to compute the collection $\col{T}$ and any analysis for the complexity of the possible combinations, i.e. translates of distinct covered point sets have not been considered in the literature.

\subsection{Our Results}\label{sec:ours}

We provide polynomial algorithms that convert the geometric cover problems, whose prototypes are planar objects of various types, to their discrete versions, i.e. the geometric set cover problems.
Each algorithm finds the reduced finite solution space that consists of so-called {\it distinct canonical translates} from infinite possible locations of translates in the plane so that the collection of the distinct canonical translates and the given point set $P$ are used to form an input instance of the corresponding geometric set cover.
The prototypes covered in this paper and the time complexities of the algorithms are listed below:
\begin{itemize}
\item for disks and their affine transforms, the geometric cover problems can be converted to the discrete versions in $O((n+k)\log n+kn)$ time,
\item for convex $m$-gons, the geometric cover problems can be converted to the discrete versions in $O((n+k)\log mn+kn)$ time,
\item for non-convex $m$-gons, the geometric cover problems can be converted to the discrete versions in $O((m^2k+mn)\log (m^2k+mn)+m^2kn)$ time,
\item for a planar object that can be decomposed into $x$-monotone Jordan curves of constant size $c$ where a pair of the curves from different translates intersect at most $s$ points, the geometric cover problem can be converted to the discrete version in $O((cn+k')\log cn+k'n)$ time or $O(cn\lambda_{s+2}(cn)+k'n)$ time,
\end{itemize}
where $n$ is the size of the input point set $P$, $k$ is the number of intersection pairs of the point inverses through $P$ (see Definition \ref{def:I}), $k'$ is the number of intersection points of the perimeters of the point inverses through $P$, $\lambda_{s+2}(n)$ is the maximum length of Davenport-Schinzel sequence of order $s+2$ for $n$ symbols.

\subsection{Related Work}\label{sec:related}

Given a set $P$ of $n$ points in the plane, the optimal placement problem is to compute a translate that contains the maximum number of points in $P$;
note that in our problem, the distinct canonical translates contain the points of $P$ in a locally maximal manner (see Definition \ref{def:CD}).
There are some optimal placement algorithms for basic planar objects:
Chazelle and Lee \cite{Cha86} presented $O(n^2)$ time algorithm for disks, and Eppstein and Erickson \cite{Epp94} proposed $O(n\log n)$ time algorithm for rectangles;
for convex polygons of $m$ vertices, Efrat et al. \cite{Efr94} devised $O(nk^*\log n\log m +m)$ time algorithm, and Barequet et al. \cite{Bar97} improved the time complexity to $O(n\log n+nk^*\log mk^* +m)$, where $k^*$ is the maximum number of points in $P$ that can be contained in a translate.

\iffalse
Agarwal et. al. - near linear time approximation algorithm
Dickerson and Scharstein - convex m-gon including rotation
\fi

\section{Main Results}\label{sec:pre}

\begin{figure}[t]
\centerline{
    \includegraphics[width=.94\columnwidth]{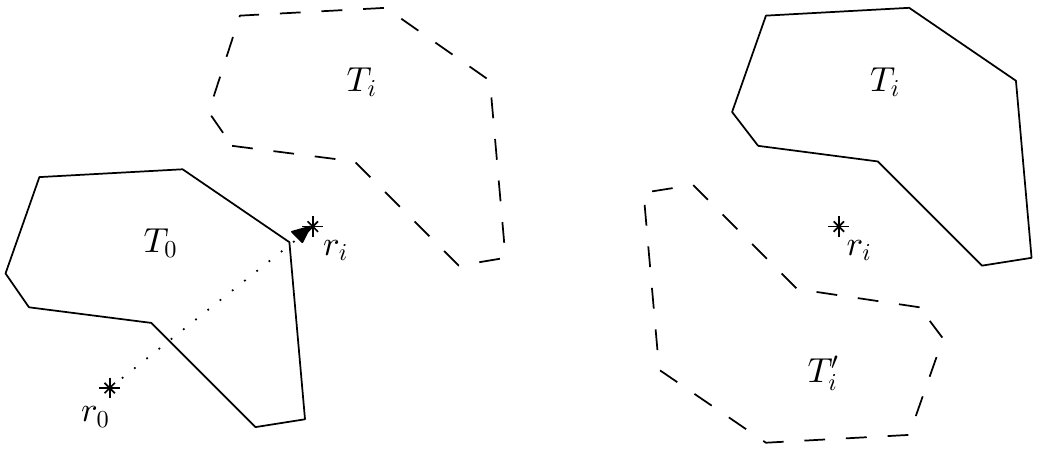}}
    \vspace*{0in}
    \caption{Left: a planar translate $T_i$ (dashed line) and its reference point $r_i$ (*) by a translation (dotted arrow) of a prototype $T_0$ (solid line). Right: the point inverse $T'_i$ (dashed line) of the translate $T_i$ (solid line) through the reference point $r_i$ (*).}
    \label{fig1}
    \vspace*{0in}
\end{figure}

In this section, we define some basic notions and derive lemmas for our discussion of the discretization of the geometric cover problem.
We begin with the following definitions on the properties and types of the translates.

\begin{defn}[Covered Point Set]
Given a set $P$ of $n$ points in the plane and a planar translate $T_i$ of an prototype $T_0$, the {\it covered point set} $P_i\subseteq P$ of $T_i$ is defined as $P_i=\{p|p\in P, p\in T_i\}$.
\end{defn}

\begin{defn}[Canonical Translate]\label{def:CD}
Given a set $P$ of points in the plane, a translate $T_i$ of $T_0$ is {\it canonical} if the covered point set $P_i\subseteq T_i$ is neither a null set nor a proper subset of $P_j$ of any translates $T_j$.
\end{defn}

\begin{defn}[Distinct Translate]
Two translates $T_i$ and $T_j$ are distinct if the covered point sets of the translates are different, i.e. $P_j\neq P_i$.
\end{defn}

\subsection{A Reduced Finite Solution Space}

Since the solution space is the 2-dimensional Euclidean plane that a planar object can be freely translated on, there can be an infinite number of translates that cover the same points in $P$.
In the set cover or the covering problems for the points, we do not need to make a distinction between the translates having the same covered point sets.
It is only required to identify and maintain the existence of translates of which the covered point sets are the same, and to save the location of a representative translate or the region of locations of those translates.
\iffalse
If there is a method to find all distinct and possible translates for a prototype of the geometric cover problem, we can obtain the discrete solution space, the collection of all the distinct and possible translates, of the geometric cover problem, which is now exactly the form of the set cover problem.
We further argue that the discretized solution space can be reduced to a smaller subspace, i.e. the collection of all distinct canonical translates.
\else
If there is a method to find all distinct representatives of possible translates for a prototype of the geometric cover problem, we can obtain the discrete solution space, the collection of all the distinct representative translates, of the geometric cover problem, which is now exactly the form of the set cover problem.
We further argue that the discretized solution space can be reduced to a smaller subspace, i.e. the collection of all distinct representatives of possible canonical translates.
\fi

\iffalse
\begin{defn}[Complete Collection of All Distinct Canonical Translates]
A collection $\col{T}_c$ is the complete collection of all distinct canonical translates if $\col{T}_c$ is the maximum cardinality collection of distinct canonical disks.
\end{defn}
\fi

\begin{lem}\label{lem:uni_can}
The union of all distinct representative canonical translates covers all the points in $P$.
\begin{proof}
For any translate $T_j$, there exists at least one canonical translate $T_i$ such that $P_j\subset P_i$, or $T_j$ is canonical since the covered point set of a canonical translate is not a proper subset of covered point sets of any translates.
Therefore, since the union of all distinct representative translates covers all the points in $P$, the union of all distinct representative canonical translates does.
\end{proof}
\end{lem}

\begin{lem}\label{lem:opt_can}
There is an optimal solution for the geometric cover problem such that all translates in the solution are canonical.
\begin{proof}
Consider one of optimal solutions for the geometric cover problem, then for each non-canonical translate $T_j$ in the solution, there is a canonical translate $T_i$ such that $P_j\subset P_i$.
Thus, we can always find an optimal solution consisting of only canonical translates by converting from an optimal solution with non-canonical translates.
\end{proof}
\end{lem}

Lemma \ref{lem:uni_can} and \ref{lem:opt_can} imply that the collection of all distinct representative canonical translates constitutes the reduced finite solution space of the geometric cover problem.

\subsection{Planar Structures Corresponding to Canonical Translates: Sink Faces}

We now introduce some definitions to present planar structures that lead to the identification on the regions of locations of the distinct (canonical) translates.

\begin{defn}[Reference Point]
The {\it reference point} $r_i$ of an object $T_i$ is an arbitrary point of which relative position to the object is invariant against planar translation.
\end{defn}

\begin{defn}[Point Inverse]
The {\it point inverse} $T'_i$ of an object $T_i$ is an affine transform created by the point inversion of $T_i$ through its reference point $r_i$.
\end{defn}

\begin{defn}[Point Inverses through $P$]\label{def:I}
The collection $\col{I}(P,T_0)$ of point inverses $T'_i$s of translates $T_i$s whose reference points are located at all the points in $P$ is called the collection of {\it point inverses through} $P$.
\end{defn}

Each element of $\col{I}(P,T_0)$, i.e. a point inverse through $p\in P$ is the set of possible positions for a translate's reference point so that the translate covers $p$.
This proposition is verified by the following finding.

\begin{find}\label{find:inv_cover}
For a point inverse $T_i'$ of a translate $T_i$ through the reference point $r_i$, a translate $T_j$ of the prototype covers $r_i$ if and only if $T_i'$ contains the reference point $r_j$ of $T_j$.
\iffalse
\begin{proof}
Consider a line $l_i$ passes through $r_i$, then $l_i\cap T_i'$ is the point inverse of $l_i\cap T_i$ through $r_i$.
Similarly, consider a line $l_j$ passes through $r_j$, i.e. the reference point of $T_j$ with the same inclination: $l_i$ and $l_j$ are parallel.
Since $T_i$ and $T_j$ are translates of each other, $l_i\cap T_i$ is a translate of $l_j\cap T_j$.
Then, $r_i$ is in $l_j\cap T_j$ if and only if $l_i\cap T'_i$ contains $r_j$.
Therefore, if $T_j$ contains $r_i$, $r_j$ is in $l\cap T_i$, where $l$ is the line passing through $r_i$ and $r_j$, and vice versa, which concludes the lemma.
\end{proof}
\fi
\end{find}

Thus, it is inferred that a common intersection of some point inverses in $\col{I}(P,T_0)$ is the set of possible positions for a translate's reference point so that the translate covers the reference points of the inverses.
Since an intersection graph of the point inverses through $P$ provides a structure of the intersections between the inverses, one may think that some property, such as maximal clique, on the intersection graph corresponds to a canonical translate:
a canonical translate corresponds to a clique in the intersection graph of the point inverses through $P$.
But, a maximal clique of the intersection graph is not necessarily matched to a canonical translate and neither is the converse true.

In this paper, we utilize an arrangement of the point inverses through $P$, which decompose the plane into points, edges, and faces, and it means that the plane is decomposed depending on the intersections of the point inverses through $P$, i.e. depending on the covered point sets of translates.

\begin{figure}[t]
\centerline{
    \includegraphics[width=.94\columnwidth]{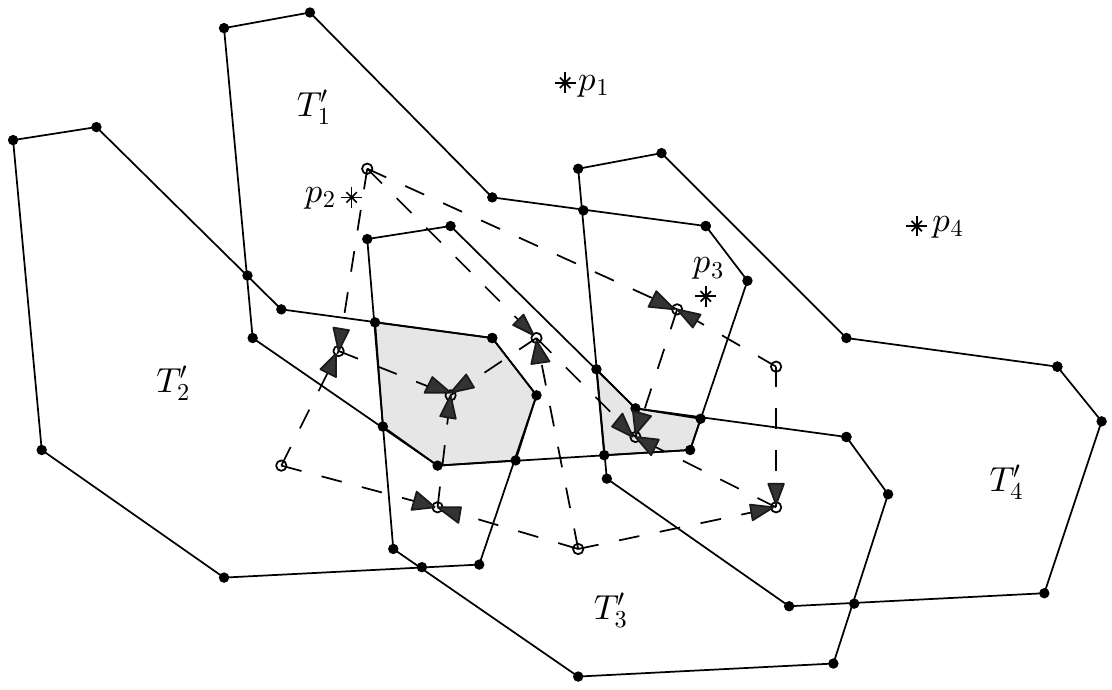}}
    \vspace*{0in}
    \caption{An example of a point inversion graph (solid), its dual graph (dashed digraph) and its sink faces (gray faces): the prototype is shown in Fig. \ref{fig1} and the points in $P$ are denoted as *. }
    \label{fig2}
    \vspace*{0in}
\end{figure}

\begin{defn}[Arrangement of Point Inverses through $P$ and Point Inversion Graph]
Given a set $P$ of $n$ points in the plane and a compact prototype $T_0$, the plane is decomposed into an {\it arrangement} $\col{A}(\col{I}(P,T_0))\equiv\col{A}(\col{I})$ of the point inverses $T'_i\in\col{I}(P,T_0)$ through $P$.
The {\it point inversion graph} $G(\col{A})$ is induced by all the vertices and edges of the arrangement $\col{A}(\col{I})$.
\end{defn}

% The point inversion graph is a planar graph.

\begin{lem}\label{lem:face}
A face of the point inversion graph $G(\col{A})$ is a connected set of points in the plane such that all translates whose reference points locate in the face have the same covered point sets,
in other words, for two translates $T_i$ and $T_j$ whose reference points locate in the face, $P_j=P_i$,
and the face is a subset of the intersection of the point inverses through $P_i$.
\begin{proof}
We consider a face as a open set, and thus it does not contain its incident edges and vertices.
Since a face of $G(\col{A})$ is a cell of the decomposed plane in the arrangement, it is a connected set of points.
Consider a face $F_i$ and the collection $\col{C}$, $\col{N}$ of inverses such that $\col{C}=\{T'_j|T'_j\in \col{I}(P,T_0),F_i\subset T'_j\}$ and $\col{N}=\{T'_j|T'_j\in \col{I}(P,T_0),F_i\cap T'_j=\phi\}$.
Note that $\col{C}\cup\col{N}=\col{I}(P,T_0)$ and the point inverses, which include edges incident to $F_i$ but do not contain $F_i$, have no intersection with $F_i$ since $F_i$ is a open set.
Thus, every translate $T_k$ whose reference point is in $F_i$, i.e. $r_k\in F_i$ has the same covered point set $P_k=\{p|p\in P,p\in\{r_j|T'_j\in\col{C}\},p\notin\{r_j|T'_j\in\col{N}\}\}$.
By the definition of $P_k$, $\col{C}$ is the collection of the point inverses through $P_k$, and thus the face $F_i$ is a subset of the intersection of the point inverses through $P_k$, the covered point set of a translate whose reference point is in $F_i$.
\end{proof}
\end{lem}

Hence, a face $F_i$ of $G(\col{A})$ is associated with a set $P_i$ of covered points by the translates whose reference points locate in the face, %: we use $P_i$ as the corresponding covered point set for the $i$th face $F_i$.
and multiple faces can be associated with the same covered point set.

Now, we present the definition of a dual graph of the point inversion graph to verify the structure of the arrangement that corresponds to the canonical translates.

\begin{defn}[Dual Graph of Point Inversion Graph]
A directed graph $H(\col{A})$ is a dual graph whose vertices correspond to the faces of the point inversion graph $G(\col{A})$ and edges correspond to the pairs of adjacent faces in $G(\col{A})$.
Each edge in $H(\col{A})$ leaves the vertex corresponding to the face of $G(\col{A})$ with the smaller covered point set in a pair of adjacent faces and enters to the other.
\end{defn}

% $H(\col{A})$ is a planar graph.

\begin{lem}\label{lem:sink}
A face of the point inversion graph $G(\col{A})$ is a connected set of points that can be the reference points of canonical translates of the same covered point set, only if the corresponding vertex in $H(\col{A})$ has no outgoing edges, i.e. the out-degree is zero;
if the compact prototype $T_0$ is convex, the face for the canonical translates of the same covered point set is unique and convex.
\begin{proof}
Let $F_i$ be a face that some canonical translates of the same covered point set have reference points on.
Then, for each pair with an adjacent face $F_j$, the covered point set $P_i$ associated with $F_i$ is exactly one larger than the covered point set $P_j$ associated with $F_j$ and $P_j\subset P_i$, because $P_i$ is of a canonical translate and thus $P_i$ is not a proper subset of any other covered point sets.
Therefore, the vertex of $H(\col{A})$ corresponding to $F_i$ has no outgoing edges.
If the prototype is convex, the face $F_i$ for canonical translates is the common intersection of the point inverses through $P_i$.
Since the inverses are convex, the common intersection is convex and obviously connected, and thus there are no locations, outside $F_i$, for the reference points of canonical disks cover $P_i$.
\end{proof}
\end{lem}

\begin{rem}[Sink Face]
Let a {\it sink face} denote a face satisfying the condition in Lemma \ref{lem:sink}.
Then, a problem finding the distinct canonical translates is equivalent to a problem searching the sink faces in the arrangement $\col{A}(\col{I})$ or the sinks of $H(\col{A})$.
Since sink nodes in a digraph can be searched by a graph traversal, the solution space of a geometric cover problem is reducible to the finite collection of the distinct canonical translates if there is a method to construct the arrangement of the point inverses through $P$.
\end{rem}

\subsection{The Number of Distinct Canonical Translates}

We now address the complexity issue on the number of distinct canonical translates for convex prototypes by investigating the number of sink faces.
Before counting the number of structures in the arrangement, we examine the number of intersections between the perimeters of two translates.
For convex planar objects, the perimeters of two translates intersect at most two times \cite{Efr94,Ked86}.

\begin{find}\label{find:cvx_intst}
Given a convex planar object $A$ and its planar translate $B$, there can be at most two intersections between the perimeters of $A$ and $B$.
\end{find}

%However, for non-convex polygons, the perimeters of two translates intersect as many times as two arbitrary (non-translate) polygons intersect.
However, for non-convex polygons, the perimeters of two translates intersect in the same order as two arbitrary (non-translate) polygons intersect: the following lemma is for the algorithm analysis in section \ref{alg_non_cvx}.

\begin{figure}[t]
\centerline{
    \includegraphics[width=.5\columnwidth]{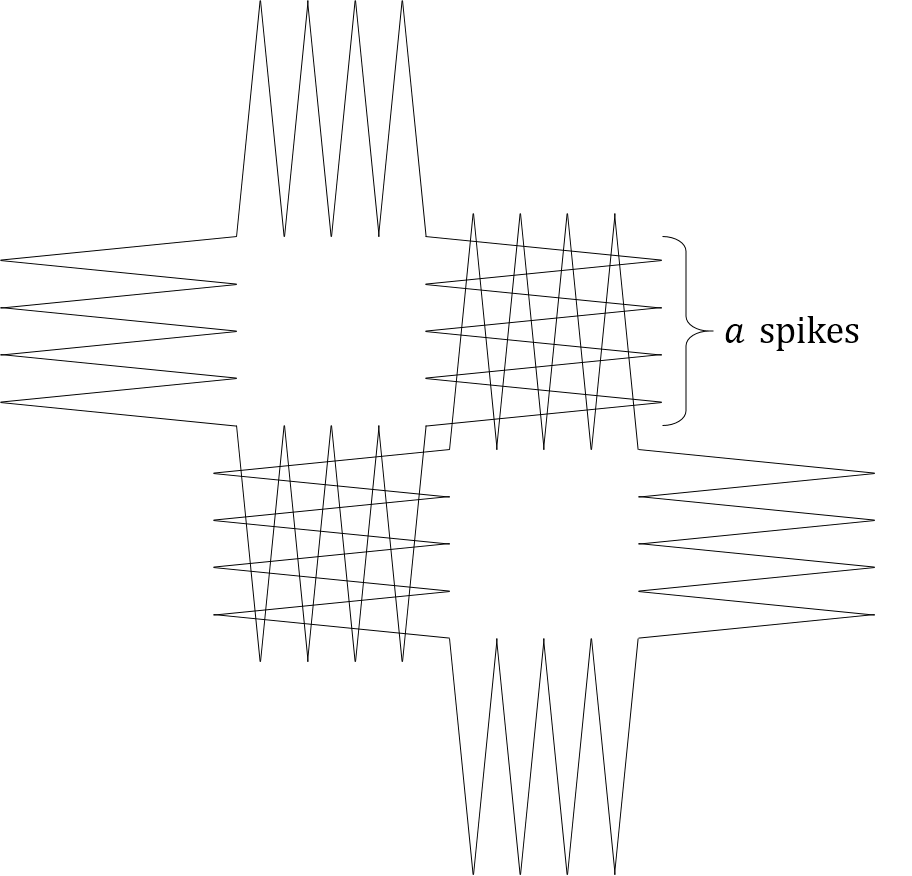}}
    \vspace*{0in}
    \caption{Two non-convex translates intersecting $\Theta(m^2)$ times: the number of edges/vertices of each translate are $m=8a$.}
    \label{fig3}
    \vspace*{0in}
\end{figure}

\begin{lem}\label{lem:non_cvx_intst}
Given a simple polygon $A$ or a polygon with holes with $m$ vertices and its planar translate $B$, there are at most $\Theta(m^2)$ intersections between the perimeters of $A$ and $B$.
\begin{proof}
Two polygons of $m$ vertices intersect at most $m(m-1)$ times, and even two translates of a prototype intersect at most  $\Theta(m^2)$ times if the prototype is a simple polygon or a polygon with holes (see Fig. \ref{fig3}).
In Fig. \ref{fig3}, each translate has $a$ spikes on each side and $m=8a$ vertices in total.
Each spike of a translate intersects with $a$ spikes of another translate and creates $4a$ intersecting vertices.
Thus, the total number of intersections of the perimeters is $4a\cdot 2a=8a^2=m^2/8=\Theta(m^2)$.
\end{proof}
\end{lem}

By Lemma \ref{lem:face}, the non-empty and distinct translates forming the discrete solution space of the geometric cover problem correspond to some faces in the point inversion graph $G(\col{A})$.
Since multiple faces can be associated with the same covered point set, the number of non-empty and distinct translates less than or equal to the number of faces in $G(\col{A})$.

\begin{lem}\label{lem:N_face}
For a convex and compact prototype $T_0$, the number of non-empty and distinct translates is less than or equal to $2k+2+e_0$, where $k$ is the number of intersection pairs of the point inverses $T'_i\in\col{I}(P,T_0)$ through $P$ and $e_0$ is the number of non-intersecting point inverses in $\col{I}(P,T_0)$.
\begin{proof}
Under the general position assumption, the point inversion graph has two types of vertices: the {\it inherent vertices} that originated from the vertices of the point inverses through $P$ and the {\it intersection vertices} that are the intersections of the perimeters of the inverses.
We modify the point inversion graph, which is a planar graph, by replacing each chain of edges and inherent vertices between two intersection vertices by two edges each of which is incident to one of the intersection vertices and an added vertex as the common endpoints of the two edges.
This modification is only for the point inverses that make any intersection with others; for the vertices and edges from non-intersecting point inverses, we just ignore them since it is clear that there are $e_0$ faces created by them.
Note that the number of faces of the graph is not changed by this modification.
Let $e_1$ denote the number of the replaced edges between the intersection vertices, then the number of edges in the modified point inversion graph is $e_1$.
By Lemma \ref{find:cvx_intst}, the number of the intersection vertices is $2k$ and the number of the added vertices by the modification is $1/2e_1$ (one vertex between two replaced edges), and thus the number of vertices in the modified point inversion graph is $2k+1/2e_1$.
From Euler's formula for a planar graph, $v-e+f=(2k+1/2e_1)-(e_1)+f=2k-1/2e_1+f=2$.
Since each intersection vertex has exactly four incident edges, $e_1=4\cdot 2k=8k$.
Thus, there are $2-2k+1/2e_1=2k+2$ faces in the point inversion graph except the faces enclosed by non-intersecting point inverses.

No three edges of the point inverses coincide under the general position assumption.
Without this assumption, if more than three edges coincide, we can always find a planar graph that has one more face than the point inversion graph by slightly moving the edges.
Therefore, the number of faces in the point inversion graph is less than or equal to $2k+2+e_0$, and by Lemma \ref{lem:face}, the faces created by the arrangement $\col{A}(\col{I})$ are the representation of possible combinations of covered point sets in the plane, and thus the number of non-empty and distinct translates is less than or equal to the number of the faces, which concludes the lemma.
\end{proof}
\end{lem}

%For a convex and compact prototype $T_0$, $k\leq n^2$.

By Lemma \ref{lem:sink}, the distinct canonical translates correspond to the sink faces of $G(\col{A})$.
In general, a sink face is not always the set of points that can be the reference point of a distinct canonical translate, and even in the case that the sink face is associated with the covered point set of a canonical translate, there can be multiple faces associated with the same covered point set.
This implies that the number of distinct canonical translates is less than or equal to the number of sink faces of $G(\col{A})$.
However, if the prototype of the geometric cover problem is convex, the number of distinct canonical translates is equal to the number of the sink faces since the sink faces are unique and convex.

\begin{lem}\label{lem:N_sink}
For a convex and compact prototype $T_0$, the number of distinct canonical translates is less than or equal to $k+e_0$.
\begin{proof}
We will bound the number of distinct canonical translates by bounding the number of sink faces of $G(\col{A})$.
Suppose that the point inverses through $P$ are introduced in the plane one by one, and then $i-1$ inverses are already exist in the plane when the $i$th inverse $T'_i$ is added.
Let $m_i$ denote the number of sink faces in the arrangement $\col{A}_i$ of $i$ point inverses.
The addition of $T'_i$ produce additional decomposition of the arrangement $\col{A}_{i-1}$ of pre-existing $i-1$ inverses.
The faces of $\col{A}_{i-1}$ totally contained in $T'_i$ are not divided by the addition whereas the faces of $\col{A}_{i-1}$ intersecting the boundary of $T'_i$ are divided.
Let the number of intersection pairs between $T'_i$ and the pre-existing $i-1$ inverses be denoted by $k_i$; $2k_i$ is the number of intersection vertices (see the proof of Lemma \ref{lem:N_face}) created by the addition of $T'_i$.
Then, $T'_i$ divides at most $2k_i$ faces of $\col{A}_{i-1}$ and create at most $k_i$ new sink faces (note that sink faces are convex if the prototype is convex).
We then obtain the following recurrence relation: $m_{i-1}+k_i\leq m_i$.
Adding all the point inverses, the number of sink faces of $G(\col{A})$ is $m_n\leq\sum_1^nk_i=k$.
Since a non-intersecting point inverse generates a sink face of $G(\col{A})$, which corresponds to a canonical translate covering a single point in $P$, the number of distinct canonical translates is less than or equal to $k+e_0$, identical to the number of the sink faces.
\end{proof}
\end{lem}

\begin{figure}[t]
\captionsetup[subfigure]{labelformat=empty}
\centering{
    \subfloat[]{
        \includegraphics[width=0.4\columnwidth]{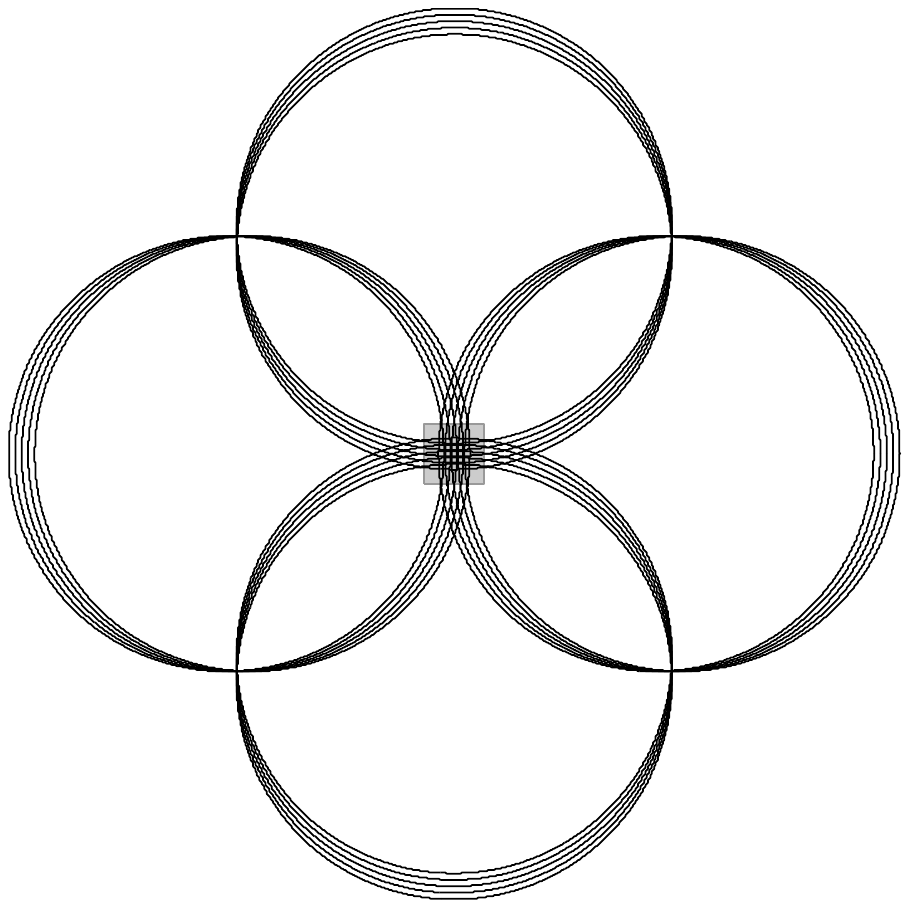}
    }
    \subfloat[]{
        \includegraphics[width=0.4\columnwidth]{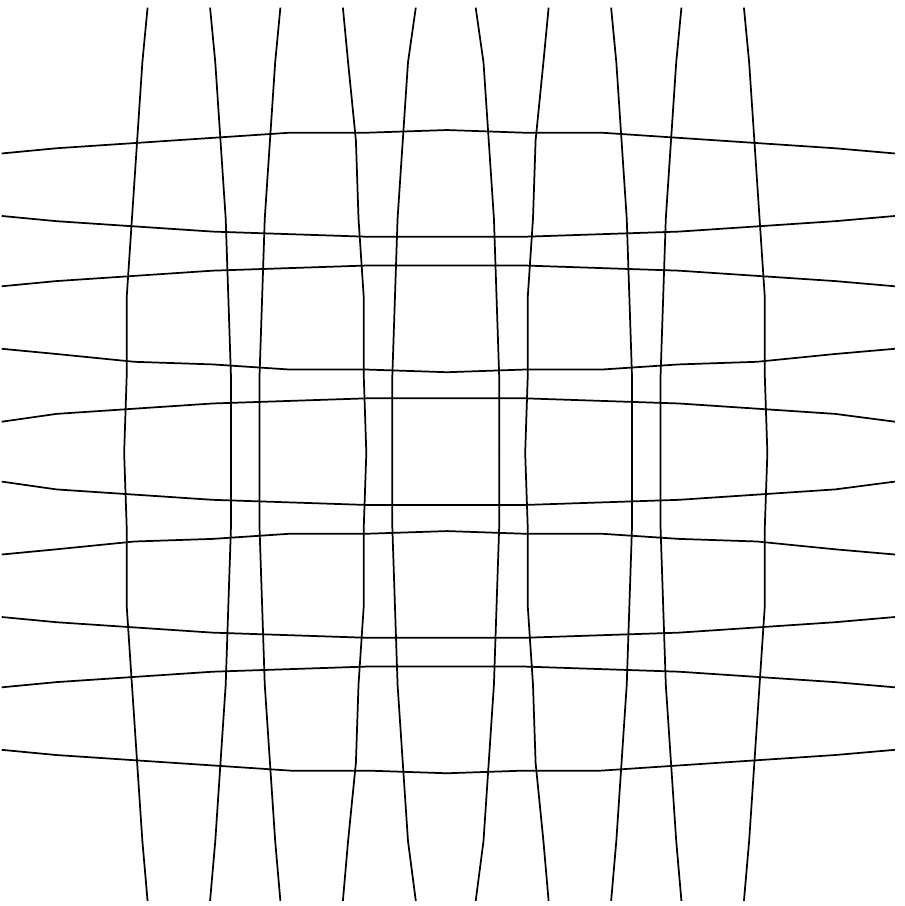}
    }}
    \vspace*{0in}
    \caption{Left: an arrangement of convex point inverses that has $\Theta(k)$ sink faces. Right: the magnified view of the gray box in the left figure.}
    \label{fig4}
    \vspace*{0in}
\end{figure}

Finally, we state about the tight bounds of the number of distinct canonical translates and the sum of the cardinality of their covered point sets.

\begin{lem}\label{lem:N_CD}
For a convex and compact prototype $T_0$, the number of distinct canonical translates is at most $\Theta(k+n)$ and the sum of the cardinalities of the covered point sets for the distinct canonical translates is at most $\Theta(kn)$.
%and for non-convex polygons with $m$ vertices, the numbers are $\Theta(m^2k)$ and $\Theta(m^2kn)$, respectively.\footnote{Note that the number $k$ of intersection pairs of the point inverses through $P$ is at most $n(n-1)=O(n^2)$.}
\begin{proof}
By Lemma \ref{lem:N_sink}, the number of distinct canonical translates for a convex prototype is $O(k+n)$ since $e_0\leq n$.
The maximum cardinality of the covered point set of a translate is trivially $n$ and the one of a canonical translate covering the reference point of a non-intersecting point inverse in $\col{I}(P,T_0)$ is 1.
Thus, the sum of the cardinalities of the covered point sets for the distinct canonical translates is $O(kn)$.
There exists a case of convex point inverses that shows these bounds are tight (see Fig. \ref{fig4}). %: thus respectively, $\Theta(k)$ and $\Theta(kn)$.
In Fig. \ref{fig4}, circles are placed densely in four directions, and let the number of the circles in each direction be $a$.
Then the total number of the circles is $n=4a$.
From the magnified view in Fig. \ref{fig4}, it is confirmed that there are $a^2=1/16n^2$ sink faces (convex faces), and the number of intersection pairs of the circles is at least $\Omega(a^2)$ since the number of intersections in the magnified view is $4a^2$.
Since the number of intersection pairs $k\leq n(n-1)$, $k=\Theta(a^2)$ and thus the number of distinct canonical disks is at most $\Theta(k)$.
Note that $2a+2=\Theta(n)$ circles contain each sink faces in the magnified view, therefore, the sum of the cardinalities of the covered point sets for the distinct canonical disks is $O(kn)$.
We can easily draw similar arrangements for other convex objects, thus the numbers concerned in this lemma are $\Theta(k)$ and $\Theta(kn)$, respectively.
\iffalse
For non-convex polygons, Lemma \ref{lem:non_cvx_intst} states that there are at most $\Theta(m^2)$ intersections between the perimeters of a pair of non-convex point inverses, which results in $\Theta(m^2k)$ vertices of the arrangement $\col{A}(\col{I})$ induced by $k$ intersection pairs.
As shown in the proof of Lemma \ref{lem:N_face}, the number of faces of a planar graph is the same order of the number of vertices, i.e. $\Theta(m^2k)$, and by Lemma \ref{lem:non_cvx_intst} and Fig. \ref{fig3}, the faces corresponding to a canonical disk appear at most $m^2$ times.
If the disks in Fig. \ref{fig4} are replaced by densely spiked polygons like in Fig. \ref{fig4}, there are $\Theta(m^2k)$ faces corresponding to a canonical disk
Therefore, the number of distinct canonical translates for a non-convex prototype is $\Theta(m^2k)$
\fi
\end{proof}
\end{lem}

\section{Algorithms}\label{sec:alg}

This section provides some algorithms to report all distinct representative canonical translates of a given geometric cover problem in the plane:
the algorithms for disks, convex and non-convex polygons, and planar objects decomposable into finite x-monotone Jordan curves are presented.
By Lemma \ref{lem:opt_can}, the collection of the distinct canonical translates constitutes a reduced finite solution space of the geometric cover problem, which is now converted to the geometric set cover.
Thus, the algorithms presented in this section are also discretization algorithms for the geometric cover problem.

\iffalse
Before the detail description and analysis of the algorithms, we state that,
by Lemma \ref{lem:opt_can} and \ref{lem:sink}, and Lemma \ref{lem:disk} to \ref{lem:jordan}, it is concluded that the geometric cover problem in the plane can be converted to the geometric set cover in polynomial time for the prototype listed in Theorem \ref{thm}.
\fi

\subsection{Algorithms for Disks}\label{sec:alg_disk}

We present two algorithms for the canonical translate reporting problem where the prototype is a disk or an affine transform of a disk:
if the prototype is an affine transform of a disk, e.g. an ellipse, the problem can be converted by transforming the point set $P$ and the prototype $T_0$ with the inverse of the affine transform.
The first algorithm is based on the plane sweep, a fundamental technique in computational geometry, over the arrangement of the point inverses through a given point set $P$.
Some proper modifications to a generic plane sweep algorithm enable reporting all distinct representative canonical translates, i.e. canonical disks, in $O((n+k)\log n)$ time and reporting the covered point sets of them in $O(kn)$ time.
The second algorithm is to traverse the point inversion graph of the arrangement.
Since the point inverse of a disk is also a disk and they are identical when the reference point is the center of the disk, the algorithm uses fixed-radius near neighbors of the points in $P$ to create the point inversion graph.
Then, the canonical disks are obtained by traversing the graph to find all convex faces, and the whole process takes the same order of times as the first algorithm for the reporting problems.

\subsubsection{Plane Sweep Algorithm for Reporting Canonical Disks}\label{sec:alg_disk_sweep}

The plane sweep algorithm for reporting canonical disks uses a vertical line sweeping over the fixed-radius disks centered on the points in $P$.
The collection of these fixed-radius disks is a simple implementation of the collection $\col{I}(P,T_0)$ of point inverses through $P$ by putting the reference points of translates (i.e. disks) at those centers.
To form an arrangement of the disks from the sweeping, the boundary of each disk is divided into an upper and a lower semicircles intersecting at their leftmost and rightmost vertices.

\null

{\bf Sweep Line.}
The sweep line moves left to right over the arrangement of the disks and intersects with the semicircles and the faces of the arrangement.
The sweep line in a generic plane sweep maintains sorted line segments or curves that intersect with the sweep line.
In our algorithm, at every interval of the sweep line divided by intersections with the semicircles, we also store the pointer to the region that is the swept part of a face where the interval belongs.
The semicircles intersecting the sweep line and intervals between them are stored into a {\it sweep line status} in sorted order.
For each region incident to the sweep line, the subset of points in $P$ covered by a disk centered at the region and the convexity of the region are maintained.
The cardinality of the covered point set is maintained instead of the set itself in case of reporting only the distinct canonical disks.
The convexity of the faces or the left parts of the faces we have swept can be determined by checking their incident arcs, i.e. some parts of semicircles since the sweep line moves left to right.

\null

{\bf Events.}
There are three types of events: leftmost vertex events, rightmost vertex events, and intersection events.
The $n$ leftmost vertex events and $n$ rightmost vertex events are computed and sorted by their x-coordinates and added to a {\it event queue} in the initialization.

For every leftmost vertex event, a {\it new} convex region starts between the upper and lower semicircles incident to the leftmost vertex, and the semicircles and the new interval between them are inserted into the sweep line status.
The pre-existing interval of the {\it old} region on the left side of the event vertex is now divided into two parts that have the same pointers to the old region which becomes non-convex.
The covered point set of the new convex region is created by adding the center of the new semicircles to the covered point set of the old region of divided intervals.

For a rightmost vertex event, the region on the left side of the vertex ends and becomes a face, and thus the corresponding interval and semicircles are deleted from the sweep line status.
If this old region is convex until it ends, it is the set of points that can be the center of a distinct canonical disk.
The neighboring intervals and corresponding regions are then merged, since they are identical in principle.

The last type of events is intersection event, where the order of intersecting semicircles in the sweep line status is changed, and one old region ends while one new region starts.
There are four possible cases of intersections between two types of semicircles: upper-upper, lower-lower, upper-lower, and lower-upper semicircles intersect.
if two upper semicircles or two lower semicircles intersect, the new region is non-convex and the convexity of other regions remain as they are.
%for the covered point set of the new region, the center of one semicircle is deleted from the covered point set of the region ended at the intersection and the center of another semicircle is added;
if an upper and a lower semicircle intersect, the region above and below the intersection is non-convex and the new region is convex only if the old region is non-convex.
The covered point set of the new region is properly modified from the covered point set of a neighboring region.

At every event of any type, the intersections between newly neighboring semicircles are computed, and if a new intersection is detected, the intersection point event is inserted into the event queue.
There is no reason to maintain the data for the non-convex regions that has no pointers in the sweep line status, and thus those regions are deleted after the events.

\begin{figure}[t]
\centerline{
    \includegraphics[width=.5\columnwidth]{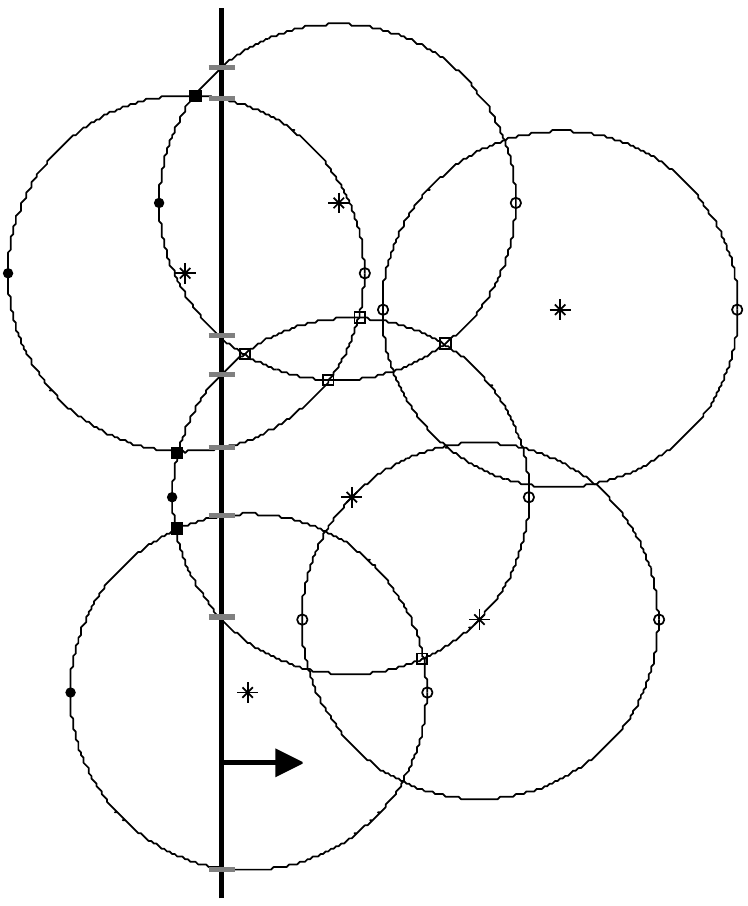}}
    \vspace*{0in}
    \caption{A graphical representation of the presented plane sweep algorithm for reporting canonical disks. The sweep line (vertical thick line) moves left to right over the arrangement of disks and is divided into intervals by the intersections with the boundaries (circles) of the disks. Leftmost and rightmost vertex events (circle dots) and calculated intersection events (square dots) are plotted: the processed events are marked as solid markers and the events in the queue are denoted by empty markers.}
    \label{fig5}
    \vspace*{0in}
\end{figure}

\null

{\bf Algorithm}
\begin{enumerate}
\item Divide the circles whose centers are the points $p\in P$ in the plane into upper and lower semicircles and insert the leftmost and rightmost vertices to the event queue $Q$ in sorted order by their x-coordinates.
\item While $Q$ is not empty, extract and process the next event from $Q$, and test and add intersections of newly neighboring semicircles after the event.
\begin{enumerate}[a)]
\item Leftmost vertex event: Create new region and divide the intervals of the old region by the semicircles incident to the event vertex. Update the convexities of the regions and create the covered point set of the new region.
\item Rightmost vertex event: Delete the interval and semicircles of the old region from the sweep line status, and delete the old region unless it is convex. The neighboring intervals and regions are merged.
\item Intersection event: Create new region and delete the interval of the old region, and delete the old region unless it is convex and corresponding intervals exist. Update the convexities of the regions incident to the intersection and create the covered point set of the new region.
\end{enumerate}
\item Report the faces for the centers of all distinct representative canonical disks and their covered point sets.
\end{enumerate}

\null

{\bf Analysis.}
For a given $n$ points in $P$, the number of semicircles are $2n$, and the number of intersection points are $2k$ where $k$ is the number of intersection pairs of circles centered on the points in $P$.
The initialization takes $O(n\log n)$ time for sorting the vertices in the event queue $Q$.
For each event, $O(\log n)$ time is required to update the event queue and the sweep line status, and other operations take constant time except for the creation of the covered point sets.
During the sweeping, the covered point set of a region is unchanged once it is created, and the sum of the cardinalities of the covered point sets of all distinct representative canonical disks is $O(kn)$, which concludes that $O(kn)$ time is required for the operations of creating the covered point sets in total.
The total number of the events is $O(n+k)$, and thus reporting all distinct representative canonical disks and their covered point sets can be done in $O((n+k)\log n+kn)$ time, and for reporting only the canonical disks, it takes $O((n+k)\log n)$ time.

\subsubsection{Traverse Algorithm for Reporting Canonical Disks}\label{sec:alg_disk_trav}

We introduce the second algorithm for reporting canonical disks that is to traverse the point inversion graph to find convex faces each of which is the set of connected points that can be the center of a distinct canonical disk.
To identify the vertices of the point inversion graph, any algorithm finding fixed-radius near neighbors of a given point set $P$ can be used.
Any two fixed-radius near neighbors are the centers of the same radius circles that intersect each other.
The intersections on each circle are sorted to build a doubly-connected-edge-list (DCEL) implementation of the point inversion graph.
We traverse the point inversion graph in clockwise (CW) direction that means every time we encounter a vertex, we always take the first edge, i.e. the first arc in counter-clockwise (CCW) order among the incident edges of the vertex.
The first edge is the rightmost edge in the view of the traversing path, thus the path is rotating CW.
Therefore, if the path arrives at any vertex of a convex face, we will definitely find the convex face and then start a new traversing path from an unvisited edge.
The visited vertices and edges never be visited again and a traversing path is ended if we find a convex face or encounter a visited vertex.
By this procedure, all edges are visited until all distinct representative canonical disks are identified.
The set of covered points needs not to be maintained for all faces incident to the edges in the path, but rather it is enough to compute the set whenever a convex face is found.

\null

{\bf Algorithm}
\begin{enumerate}
\item Find all intersection pairs of the circles whose centers are the points $p\in P$ in the plane by any algorithm finding fixed-radius near neighbors of $p\in P$.
\item For each circle, sort the intersection points with other circles and link them CW (right side is interior) and CCW (left side is interior). Create and save the edges for each intersection to form a DCEL implementation of the arrangement of the circles. Mark all vertices and edges in the arrangement as {\it new} ones.
\item Select an arbitrary edge in the arrangement to start a traversing path. % and compute and create the set of covered points by the circle centered on the edge.
\item Traversing the arrangement CW, every time a vertex is encountered, mark the previous edge as an {\it old} (traversed) edge.
\begin{enumerate}[a)]
\item If the encountered vertex is new, mark it as old and select the first edge in CCW order among the incident new edges. %Add the center of the circle, which includes the selected edge, to the covered point set.
\item If an old vertex is encountered, then create a canonical disk centered on the encircled convex face by the traversing path and mark all old vertices and edges as {\it obsolete} ones.
\item If an obsolete vertex is reached, make all old vertices and edges obsolete.
\end{enumerate}
\item Until all edges are obsolete or all vertices are obsolete, select a new edge and then return to step 4.
%\item Until all edges are obsolete or all vertices are obsolete, select a new edge {\red incident to an obsolete vertex}, and compute and create the set of covered points by the circle centered on the edge, and then return to step 4.
\item Compute the covered point sets of the created canonical disks.
\item Report all distinct representative canonical disks and their covered point sets.
\end{enumerate}

\null

{\bf Analysis.}
The all fixed-radius near neighbors of a given $n$ points can be computed in $O(n+k)$ \cite{Ben77}.
Sorting intersections of each circle requires $O(k_i\log n)$ time where $k_i$ is the number of intersections of the circle, and thus the process takes $O(\sum_i^n k_i\log n)=O(k\log n)$ time in total for all circles.
Building the point inversion graph takes $O(n+k)$ time and graph traversing requires the same order of time.
Therefore, reporting all distinct representative canonical disks can be done in $O((n+k)\log n)$ time and their cover point sets can be computed in $O(kn)$ time.

\begin{lem}\label{lem:disk}
If the prototype $T_0$ of the geometric cover problem is a disk or an affine transform of a disk, there are some algorithms that can report all distinct representative canonical translates in $O((n+k)\log n)$ time, and report the covered point sets of all the canonical translates in $O((n+k)\log n+kn)$ time.
\end{lem}

\subsection{Algorithm for Convex Polygons}\label{sec:alg_convex}

For the canonical translate reporting problem where the prototype is a convex polygon, we use the plane sweep technique as it is used for the problem of disks.
If the prototype is a convex polygon of $m$ vertices, the inverses of the translates whose reference points are the points in a given set $P$ are also convex polygons of $m$ vertices.
We rotate the axes of the plane so that no edges of the polygons are parallel to the vertical axis of the plane, and then divide each inverse polygon into an upper and a lower polygonal chain at the leftmost and rightmost vertices of the inverse polygon.
The sweep line status and the event queue in the algorithm stores almost the same entities as in \ref{sec:alg_disk_sweep}, but instead of upper/lower semicircles, the sweep line status maintains the upper/lower polygonal chains, and the vertices of the chains except the end-points (the leftmost and rightmost vertices of the inverse polygons) do not trigger any events.
If the form of report and identification of a distinct canonical translate consists of a sample reference point of the canonical translate, rather than the face for the canonical translate, and its covered point set, we do not need to construct the arrangement and thus not need to maintain all coordinates of edges of the $2n$ polygonal chains.
It is only required to maintain the sorted list of the polygonal chains intersecting the sweep line and the intervals between the chains, and to test the intersections of neighboring chains for creating intersection events.
Since all the segments of the chains between the intersections are parts of a convex polygon, the convexity of a region is not changed in-between the intersections during the plane sweep.
Therefore, the existence of a convex face in the arrangement, which is not actually constructed, can be verified at the rightmost vertex events or intersection events.% if the {\red closing} region is convex.

Using the similar technique for convex polygons, Barequet et al. \cite{Bar97} presented an algorithm for computing an optimal translate that covers the maximum number of points in a given point set.
They made use of the same finding as in Finding \ref{find:inv_cover} to find the optimal translate by computing a location of maximum depth in the arrangement of the point inverses through $P$.
Efrat et al. \cite{Efr94} also utilized the same approach as a sub-routine for computing the smallest homothetic copy of a given polygon containing some fixed number of points.

\null

{\bf Algorithm}
\begin{enumerate}
\item Rotate the axes of the plane so that no edges of the prototype are parallel to the vertical axis. Divide the inverses whose reference points are the points $p\in P$ into x-monotone polygonal chains and insert the leftmost and rightmost vertices to the event queue $Q$ in sorted order by their x-coordinates.
\item While $Q$ is not empty, extract and process the next event from $Q$, and test and add intersections of newly neighboring polygonal chains after the event.
\begin{enumerate}[a)]
\item Leftmost vertex event: Create new region and divide the intervals of the old region by the polygonal chains incident to the event vertex. Update the convexities of the regions and create the covered point set of the new region.
\item Rightmost vertex event: Delete the interval and polygonal chains of the old region from the sweep line status, and delete the old region unless it is convex. The neighboring intervals and regions are merged.
\item Intersection event: Create new region and delete the intervals of the old region, and delete the old region unless it is convex and corresponding intervals exist. Update the convexities of the regions incident to the intersection and create the covered point set of the new region.
\end{enumerate}
\item Report all distinct representative canonical translates and their covered point sets.
\end{enumerate}

\null

{\bf Analysis.}
For a given $n$ points in $P$, by Finding \ref{find:cvx_intst}, the arrangment of point inverses through $P$ has at most $2k$ vertices from the intersections of the polygonal chains, and thus $O(n+k)$ events are processed during the sweep algorithm.
The algorithm is basically the same as the plane sweep algorithm for canonical disk reporting except the test to find intersections between two neighboring polygonal chains.
The brute-force approach for the test requires $O(m)$ time for every event, but there are efficient methods that run in $O(\log m)$ time, one of them is based on binary search between two convex polygonal chains \cite{Dob83}.
Thus, the time complexity of the algorithm is $O((n+k)(\log n + \log m)+kn)=O((n+k)(\log nm)+kn)$.

\null

\begin{lem}\label{lem:cvx}
If the prototype $T_0$ of the geometric cover problem is a convex $m$-gon, there is some algorithm that can report all distinct representative canonical translates in $O((n+k)\log mn)$ time, and report the covered point sets of all the distinct representative canonical translates in $O((n+k)\log mn+kn)$ time.
\end{lem}

\subsection{Generalization}

The algorithms presented in section \ref{sec:alg_disk} and \ref{sec:alg_convex} are to search and report the distinct canonical translates and the covered point sets by traversing the point inversion graph or sweeping the arrangement of point inverses through $P$.
This approach can be generalized to non-convex polygons and more general types of planar objects: the objects decomposable into a finite number of Jordan curves.

\subsubsection{Simple Polygons or Polygons with Holes}\label{alg_non_cvx}

For non-convex simple polygons or polygons with holes, we also use the plane sweep technique;
there are two differences in the algorithm for these types of polygons from the one for convex polygons.
The first point is because a non-convex polygon of $m$ vertices is no more divisible into two x-monotone and convex polygonal chains, but rather it is decomposable to $O(m)$ structures of those properties.
Thus, we simply regard the arrangement of $n$ polygons of $m$ vertices as the one of $nm$ line segments, and for every vertex encountered during sweeping, except the leftmost and rightmost ones of polygons, we replace the line segment in the sweep line status with the next line segment incident to the vertex.
The second difference arises from the fact that a sink face in the arrangement of non-convex polygons is neither unique and nor convex.
Thus, a canonical translate can not be identified by the convexity test of a face, but instead, can be identified by checking the out-degree of a vertex in the dual graph $H(\col{A})$ of the point inversion graph;
at a leftmost vertex event or an intersection event, we update the out-degree of a vertex in $H(\col{A})$, corresponding to a face in $G(\col{A})$, by checking the interior of the line segments between the regions incident to the event vertex or intersection.
Since the axes of the plane are rotated so that no line segments, i.e. edges of the polygons are vertical, we can save which region is interior of a polygon between the regions above and below a line segment in the initialization stage.

The time complexity of the algorithm is straightforward.
We have $mn$ line segments and at most $m^2k$ intersections by Lemma \ref{lem:non_cvx_intst}, the canonical translates can be identified in $O((m^2k+mn)\log (m^2k+mn))$ time with constant time operations in each event.
%By Euler's formula for planar graphs, the number of faces is the same order of the number of vertices of the arrangement,
Since an intersection creates at most a constant number of faces, the number of faces is the same order of or less than the maximum number of vertices in the arrangement, and thus the covered point sets of the canonical translates are obtained in additional $O(m^2k\cdot n)$ time by computing the covered point sets of all regions intersecting the sweep line.

\begin{lem}\label{lem:non_cvx}
If the prototype $T_0$ of the geometric cover problem is a simple polygon or a polygon with holes of $m$ vertices, there is some algorithm that can report all distinct representative canonical translates in $O((m^2k+mn)\log (m^2k+mn))$ time, and report the covered point sets of all the distinct representative canonical translates in $O((m^2k+mn)\log (m^2k+mn)+m^2kn)$ time.
\end{lem}

\subsubsection{$x$-monotone Jordan Curves}

If a planar object is decomposable into some Jordan curves where the operations for the curves can be processed at unit costs, we can deal with the canonical translates of the object in a practical time complexity;
Since the canonical translates and their covered point sets can be computed by traversing the dual graph $H(\col{A})$ of the point inversion graph, any algorithm for the arrangement of the planar objects is utilizable for reporting the canonical translates and the covered point sets.
Edelsbrunner et al. \cite{Ede92} presented an algorithm for constructing the arrangement of $n$ $x$-monotone Jordan curves each pair of which intersect at most $s$ points in $O(n\lambda_{s+2}(n))$ time and $O(n^2)$ space by the vertical decomposition of the arrangement: $\lambda_{s+2}(n)$ is the maximum length of Davenport-Schinzel sequence of order $s+2$ for $n$ symbols.
If the prototype of the geometric cover problem is decomposable into $x$-monotone Jordan curves of constant size $c$ where a pair of the curves from different translates intersect at most $s$ points, the arrangement can be obtained, less tightly, in $O(cn\lambda_{s+2}(cn))$ time.
Let $k'$ denote the number of intersections of the perimeters of the point inverses through $P$, which means the number of vertices of the arrangement is $k'+2cn$ assuming the Jordan curves originally have vertices only at both ends, and there are $O(k')$ faces in the arrangement.
Since the maximum cardinality of a covered point set is $n$, reporting the canonical translates and their covered point sets can be done by traversing the dual graph $H(\col{A})$ in $O(k'n)$ time: the covered point set for a face (a vertex in $H(\col{A})$)  is computed by utilizing the one of a neighboring face where adding new entry requires unit cost and deletion takes at most $n$ times.
On the other hand, the plane sweep technique is still valid for this case: the worst case running time is worse than the former.
With the unit cost assumption for the operations of the Jordan curves, the arrangement of the point inverses through $P$ is constructed in $O((cn+k')\log cn)$ time.
As the same approach traversing the dual graph $H(\col{A})$, the canonical translates and their covered point sets can be computed in $O(k'n)$ time.

\begin{lem}\label{lem:jordan}
If the prototype $T_0$ of the geometric cover problem can be decomposed into $x$-monotone Jordan curves of constant size $c$, each pair of which intersect at most $s$ points, there are some algorithms that can report all distinct representative canonical translates and their covered point sets in $O((cn+k')\log cn+k'n)$ time or $O(cn\lambda_{s+2}(cn)+k'n)$ time.
\end{lem}

\iffalse
\begin{conj}
For a convex and compact planar object $T_0$, by topological sweeping, reporting canonical translates is processed in $O(n\log n+k)$ time, and reporting $P_i$s is processed in $O(n\log n+kn)$ time; it is optimal algorithm.
\end{conj}
\fi

\subsection{Summary}

By Lemma \ref{lem:opt_can} and \ref{lem:sink}, and Lemma \ref{lem:disk} to \ref{lem:jordan}, it is concluded that the geometric cover problem in the plane can be converted to the geometric set cover in polynomial time for the prototype listed in Theorem \ref{thm}.

\begin{thm}\label{thm}
If the prototype of the geometric cover problem belongs to one of the types listed below, the geometric cover problem is polynomial reducible to the geometric set cover:
\begin{itemize}
\item an affine transform of a disk,
\item a polygon,
\item a planar object decomposable into finite $x$-monotone Jordan curves each pair of which intersect at most constant times.
\end{itemize}
\end{thm}

\subsection{Remarks}

The approach proposed in this section can be generalized to the geometric cover problem in a higher dimension.
If there exists some efficient algorithm for constructing the arrangement of the point inverses for a given prototype and a point set $P$ in $d$-dimension, $d>2$, the reduced finite solution space for defining the geometric set cover can be obtained.

By incremental construction, an arrangement of $n$ hyperplanes in $\dom{R}^d$ can be constructed in $\Theta(n^d)$ time \cite{Ede86}, which means that for any polytopes described by combinations of hyperplanes, the arrangement can also be constructed but with a redundant computation induced by the arrangement of the hyperplanes.
For $n$ convex polytopes with a total of $m$ vertices, the complexity of the arrangement of the convex polytopes is known to be $\Theta(m^{\lfloor d/2\rfloor}n^{\lceil d/2\rceil})$ \cite{Aro91}, and there are $O(n^d)$ maximally covered cells, an extended concept of the sink faces, in the arrangement \cite{Gui98}.
Since $m>n$, an optimal algorithm finding the maximally covered cells can be an efficient method for the discretization of the geometric cover problem in a high dimensional space, and it can be shown that sample points of all candidate maximally covered cells, not necessarily the maximally covered cells, can be found in $O(n^dm)$ time by a slight modification of a subroutine of the algorithm presented by Guibas et al. \cite{Gui98}.

\section{Conclusion}\label{sec:res}

This paper showed that the geometric cover problem in the plane is reducible to its discrete version, the geometric set cover, by identifying the reduced finite solution space consisting of the distinct canonical translates among the infinite possibilities of the original solution space.
The algorithms for the conversion of the problem with some types of planar objects were also presented, which is to report the distinct canonical translates and their covered point sets.

\section*{Acknowledgements}

This work was supported by Basic Science Research Program through the National Research Foundation of Korea (NRF) and by the KI Project via KAIST Institute for Design of Complex Systems.
%funded by the Ministry of Science, ICT and Future Plannning (2013008693).

\section*{References}

%\nocite{Ach13,Bar97,Bro95,Cal04,Car07,Cha86,Cla07,Cla10,Das11,Efr94,Epp94,Erl10,Fow81,Fra12,Gon91,Hoc85,Joh82,Lau08,Lia10,Mus09,Nar06}

\bibliographystyle{elsarticle-harv}
\bibliography{CGTA_Discrtz_GeoCover}

\begin{thebibliography}{30}
\expandafter\ifx\csname natexlab\endcsname\relax\def\natexlab#1{#1}\fi
\expandafter\ifx\csname url\endcsname\relax
  \def\url#1{\texttt{#1}}\fi
\expandafter\ifx\csname urlprefix\endcsname\relax\def\urlprefix{URL }\fi

\bibitem[{Acharyya et~al.(2013)Acharyya, Manjanna, and Das}]{Ach13}
Acharyya, R., Manjanna, B., Das, G.~K., 2013. Unit disk cover problem in {2D}.
  In: Computational Science and Its Applications, ICCSA 2013. Vol. 7972 of
  Lecture Notes in Computer Science. Springer Berlin Heidelberg, pp. 73--85.

\bibitem[{Aronov et~al.(1991)Aronov, Bern, and Eppstein}]{Aro91}
Aronov, B., Bern, M., Eppstein, D., 1991. Arrangements of polytopes,
  manuscript.

\bibitem[{Barequet et~al.(1997)Barequet, Dickerson, and Pau}]{Bar97}
Barequet, G., Dickerson, M., Pau, P., 1997. Translating a convex polygon to
  contain a maximum number of points. Computational Geometry 8~(4), 167--179.

\bibitem[{Bentley et~al.(1977)Bentley, Stanat, and Williams~Jr}]{Ben77}
Bentley, J.~L., Stanat, D.~F., Williams~Jr, E.~H., 1977. The complexity of
  finding fixed-radius near neighbors. Information Processing Letters 6~(6),
  209--212.

\bibitem[{Br\"{o}nnimann and Goodrich(1995)}]{Bro95}
Br\"{o}nnimann, H., Goodrich, M., 1995. Almost optimal set covers in finite
  {VC}-dimension. Discrete \& Computational Geometry 14~(1), 463--479.

\bibitem[{Carmi et~al.(2007)Carmi, Katz, and Lev-Tov}]{Car07}
Carmi, P., Katz, M., Lev-Tov, N., 2007. Covering points by unit disks of fixed
  location. In: Algorithms and Computation. Vol. 4835 of Lecture Notes in
  Computer Science. Springer Berlin Heidelberg, pp. 644--655.

\bibitem[{Chazelle and Lee(1986)}]{Cha86}
Chazelle, B., Lee, D., 1986. On a circle placement problem. Computing 36~(1-2),
  1--16.

\bibitem[{Clarkson and Varadarajan(2007)}]{Cla07}
Clarkson, K.~L., Varadarajan, K., 2007. Improved approximation algorithms for
  geometric set cover. Discrete \& Computational Geometry 37~(1), 43--58.

\bibitem[{Claude et~al.(2010)Claude, Das, Dorrigiv, Durocher, Fraser,
  L\'{o}Pez-Ortiz, Nickerson, and Salinger}]{Cla10}
Claude, F., Das, G.~K., Dorrigiv, R., Durocher, S., Fraser, R.,
  L\'{o}Pez-Ortiz, A., Nickerson, B., Salinger, A., 2010. An improved
  line-separable algorithm for discrete unit disk cover. Discrete Mathematics,
  Algorithms and Applications 2~(1), 77--87.

\bibitem[{C\u{a}linescu et~al.(2004)C\u{a}linescu, M\u{a}ndoiu, Wan, and
  Zelikovsky}]{Cal04}
C\u{a}linescu, G., M\u{a}ndoiu, I., Wan, P.-J., Zelikovsky, A.~Z., 2004.
  Selecting forwarding neighbors in wireless ad hoc networks. Mobile Networks
  and Applications 9~(2), 101--111.

\bibitem[{Das et~al.(2011)Das, Fraser, L\'{o}pez-Ortiz, and Nickerson}]{Das11}
Das, G., Fraser, R., L\'{o}pez-Ortiz, A., Nickerson, B.~G., 2011. On the
  discrete unit disk cover problem. In: WALCOM: Algorithms and Computation.
  Vol. 6552 of Lecture Notes in Computer Science. Springer Berlin Heidelberg,
  pp. 146--157.

\bibitem[{Dobkin and Kirkpatrick(1983)}]{Dob83}
Dobkin, D.~P., Kirkpatrick, D.~G., 1983. Fast detection of polyhedral
  intersection. Theoretical Computer Science 27~(3), 241 -- 253, special Issue
  Ninth International Colloquium on Automata, Languages and Programming (ICALP)
  Aarhus, Summer 1982.

\bibitem[{Edelsbrunner et~al.(1992)Edelsbrunner, Guibas, Pach, Pollack, Seidel,
  and Sharir}]{Ede92}
Edelsbrunner, H., Guibas, L., Pach, J., Pollack, R., Seidel, R., Sharir, M.,
  1992. Arrangements of curves in the plane: {T}opology, combinatorics, and
  algorithms. Theoretical Computer Science 92~(2), 319--336.

\bibitem[{Edelsbrunner et~al.(1986)Edelsbrunner, O'Rourke, and Seidel}]{Ede86}
Edelsbrunner, H., O'Rourke, J., Seidel, R., 1986. Constructing arrangements of
  lines and hyperplanes with applications. SIAM Journal on Computing 15~(2),
  341--363.

\bibitem[{Efrat et~al.(1994)Efrat, Sharir, and Ziv}]{Efr94}
Efrat, A., Sharir, M., Ziv, A., 1994. Computing the smallest k-enclosing circle
  and related problems. Computational Geometry 4~(3), 119--136.

\bibitem[{Eppstein and Erickson(1994)}]{Epp94}
Eppstein, D., Erickson, J., 1994. Iterated nearest neighbors and finding
  minimal polytopes. Discrete \& Computational Geometry 11~(1), 321--350.

\bibitem[{Fowler et~al.(1981)Fowler, Paterson, and Tanimoto}]{Fow81}
Fowler, R.~J., Paterson, M.~S., Tanimoto, S.~L., 1981. Optimal packing and
  covering in the plane are {NP}-complete. Information Processing Letters
  12~(3), 133--137.

\bibitem[{Fraser and L\'{o}pez-Ortiz(2012)}]{Fra12}
Fraser, R., L\'{o}pez-Ortiz, A., 2012. The within-strip discrete unit disk
  cover problem. In: 24th Canadian Conference on Computational Geometry. pp.
  61--66.

\bibitem[{Gonzalez(1991)}]{Gon91}
Gonzalez, T.~F., 1991. Covering a set of points in multidimensional space.
  Information Processing Letters 40~(4), 181--188.

\bibitem[{Guibas et~al.(1998)Guibas, Halperin, Hirukawa, Latombe, and
  Wilson}]{Gui98}
Guibas, L.~J., Halperin, D., Hirukawa, H., Latombe, J.-C., Wilson, R.~H., 1998.
  Polyhedral assembly partitioning using maximally covered cells in
  arrangements of convex polytopes. International Journal of Computational
  Geometry \& Applications 8~(2), 179--199.

\bibitem[{Haussler and Welzl(1987)}]{Hau87}
Haussler, D., Welzl, E., 1987. {$\epsilon$}-nets and simplex range queries.
  Discrete \& Computational Geometry 2~(1), 127--151.

\bibitem[{Hochbaum and Maass(1985)}]{Hoc85}
Hochbaum, D.~S., Maass, W., 1985. Approximation schemes for covering and
  packing problems in image processing and {VLSI}. J. ACM 32, 130--136.

\bibitem[{Johnson(1982)}]{Joh82}
Johnson, D.~S., 1982. The {NP}-completeness column: An ongoing gulde. Journal
  of Algorithms 3~(2), 182--195.

\bibitem[{Kedem et~al.(1986)Kedem, Livne, Pach, and Sharir}]{Ked86}
Kedem, K., Livne, R., Pach, J., Sharir, M., 1986. On the union of jordan
  regions and collision-free translational motion amidst polygonal obstacles.
  Discrete \& Computational Geometry 1~(1), 59--71.

\bibitem[{Laue(2008)}]{Lau08}
Laue, S., 2008. Geometric set cover and hitting sets for polytopes in {$R^3$}.
  In: Albers, S., Weil, P. (Eds.), 25th International Symposium on Theoretical
  Aspects of Computer Science. Vol.~1 of Leibniz International Proceedings in
  Informatics (LIPIcs). pp. 479--490.

\bibitem[{Liao and Hu(2010)}]{Lia10}
Liao, C., Hu, S., 2010. Polynomial time approximation schemes for minimum disk
  cover problems. Journal of Combinatorial Optimization 20~(4), 399--412.

\bibitem[{Megiddo and Supowit(1984)}]{Meg84}
Megiddo, N., Supowit, K., 1984. On the complexity of some common geometric
  location problems. SIAM Journal on Computing 13~(1), 182--196.

\bibitem[{Mustafa and Ray(2009)}]{Mus09}
Mustafa, N.~H., Ray, S., 2009. {PTAS} for geometric hitting set problems via
  local search. In: Proceedings of the Twenty-fifth Annual Symposium on
  Computational Geometry. SCG '09. pp. 17--22.

\bibitem[{Narayanappa and Vojt\v{e}chovsk\'{y}(2006)}]{Nar06}
Narayanappa, S., Vojt\v{e}chovsk\'{y}, P., 2006. An improved approximation
  factor for the unit disk covering problem. In: 18th Canadian Conference on
  Computational Geometry. pp. 15--18.

\bibitem[{Raz and Safra(1997)}]{Raz97}
Raz, R., Safra, S., 1997. A sub-constant error-probability low-degree test, and
  a sub-constant error-probability {PCP} characterization of {NP}. In:
  Proceedings of the Twenty-ninth Annual ACM Symposium on Theory of Computing.
  STOC '97. ACM, New York, NY, USA, pp. 475--484.

\end{thebibliography}

\end{document}